\documentclass[pre,twocolumn,aps,eqsecnum]{revtex4}
\usepackage{epsfig}

\begin{document}

\title{Thermodynamic theory of dislocation-enabled plasticity}

\author{J.S. Langer}
\affiliation{Department of Physics, University of California, Santa Barbara, CA  93106-9530}

\date{\today}

\begin{abstract}
The thermodynamic theory of dislocation-enabled plasticity is based on two unconventional hypotheses. The first of these is that a system of dislocations, driven by external forces and irreversibly exchanging heat with its environment, must be characterized by a thermodynamically defined effective temperature that is not the same as the ordinary temperature.  The second hypothesis is that the overwhelmingly dominant mechanism controlling plastic deformation is thermally activated depinning of entangled pairs of dislocations.  This paper consists of a systematic reformulation of this theory followed by examples of its use in analyses of experimentally observed phenomena including strain hardening, grain-size (Hall-Petch) effects, yielding transitions, and adiabatic shear banding.  
\end{abstract}
\maketitle
\section {Introduction}

In 2010, Eran Bouchbinder, Turab Lookman and I published a paper (LBL) \cite{LBL-10} with almost the same title as this one.  In that paper, we used statistical ideas that had emerged in theories of amorphous plasticity to construct a theory of dislocations in deforming polycrystalline solids.  More recently, I have used this thermodynamic dislocation theory in analyses of a variety of experimentally observed phenomena including the dependence of strain hardening on strain rate and temperature, grain-size (Hall-Petch) effects, yielding transitions, adiabatic shear banding, and the interplay between thermal and mechanical effects in determining responses to varying histories of deformation. \cite{JSL-15,JSL-15y,JSL-16,JSL-17,LTL-17}  

The ways in which I understand the thermodynamic theory have evolved in recent years, and some of my original presentations need to be improved and put into proper perspective.  My purposes here are to clarify the logical structure of the theory, to summarize its main accomplishments so far, and to point out its limitations and the directions in which it needs to be extended. 

The thermodynamic dislocation theory is qualitatively different from other theories in this field, both in its first-principles starting point and its predictive capabilities.  The most common conventional theories either are descriptions of single dislocations, and thus do not touch on central issues such as strain hardening; or else they are attempts to construct empirical models of dislocation driven plastic deformation based directly on experimental observations. Neither of these approaches has proven to be as useful in applied materials science as, for example, quantum theories of electronic properties or statistical theories of pattern formation.  The thermodynamic theory of dislocation-enabled plasticity is an attempt to bring this central part of materials science up to a level of applicability comparable to those other parts of the field.  

The thermodynamic theory starts with two unconventional assumptions. The first of these is that a system of dislocations, driven by external forces and irreversibly exchanging heat with its environment, must be characterized by a thermodynamically well defined effective temperature that coexists with the ordinary temperature but is not the same as it.  In my opinion, there is nothing speculative about this assumption; I believe that it follows from the basic principles of nonequilibrium statistical thermodynamics.  

Second, the thermodynamic theory is based on a special physical assumption -- that the overwhelmingly dominant mechanism controlling plastic deformation is thermally activated depinning of entangled pairs of dislocations.  So far as I know, no other investigators have explored such an assumption.  Most commonly, the proposed empirical relations between stress, strain, strain rate, temperature, etc. are multi-parameter  fits to experimental data, based loosely on pictures of independent dislocations impeded by various kinds of obstacles as they move through a lattice.  These empirical fits do reveal the importance of an activation mechanism, but provide no specifics about what that mechanism might be. In contrast, the depinning model contains just two physics-based parameters: the pinning energy and a dimensionless ratio of two length scales that determines how that energy barrier is reduced by a stress field.  There are limitations to the range of validity of this model, to be discussed later in this paper; but, within that range, the depinning model has turned out to be remarkably successful.   

The next two sections of this paper, Secs.~\ref{Teff} and \ref{Depin},  are devoted, respectively, to the effective temperature and the depinning model.  In both of these sections, the analysis pertains only to steady-state deformations, where the nonequilibrium aspects of the problem are relatively easy to understand.  Nonequilibrium thermodynamics has been a difficult and controversial subject for at least a century.  In 2009, Bouchbinder and I wrote a series of papers about this topic largely because we needed a basic understanding of it in order to solve problems in amorphous plasticity.\cite{BLI-09}  I repeat the necessary parts of that analysis at the beginning of Sec.~\ref{EOM} in preparation for a systematic derivation of the equations of motion for the thermodynamic dislocation theory.  Then, in the following sections, I summarize applications of this theory, discuss its limitations, and point to what I believe are important open questions.  

\section{Elementary degrees of freedom and the Effective Temperature}
\label{Teff}

The thermodynamic dislocation theory starts with the assertion that a plastically deforming polycrystalline solid must be described by two distinct sets of elementary degrees of freedom, distinguished most importantly by the time scales on which they move. One set of these degrees of freedom consists of the configurational coordinates that determine the mechanically stable positions of all the atoms, including the positions of the dislocations. These configurations change slowly during macroscopic plastic deformation.  The other set of degrees of freedom  consists of the fast kinetic-vibrational variables that describe small fluctuations about the stable configurations.  The average energy of these fluctuations is proportional to the ordinary temperature $T$. Their motions occur on microscopic time scales, of the order of $10^{-10} s$ or less, very much smaller than the time scales ordinarily associated with mechanically driven deformation.

To a good approximation, the configurational and the kinetic-vibrational degrees of freedom describe a pair of weakly coupled subsytems of the deforming system as a whole.  Operationally, we can think of these subsystems as if they were two separate entities, with different temperatures and subject to different external forces, connected to each other only by a poor heat conductor.  These subsystems do exchange energy when groups of atoms undergo irreversible rearrangements.  The configurational subsystem briefly acquires energy from the kinetic-vibrational subsystem, and returns about the same amount of energy to it when the atoms fall into new stable positions.  These rearrangements occur on microscopic time scales but, unless the solid is near its melting point, they are rare events.  Thus, on average, the configurational motions are  macroscopically slow.  

To explore this picture without going into unnecessary detail too early in the presentation, it is useful to think of a slab of material lying in the plane of an applied shear stress, undergoing only steady-state deformation, and to focus only on the dislocations.  The dislocation lines oriented perpendicular to this plane are driven by the stress to move through a forest of entangling dislocations lying primarily in the plane, thus producing shear flow. Let the area of this slab be $A$, and let its thickness be a characteristic dislocation length, say $L$.  Denote the configurational energy and entropy of the slab by $U_0(\rho)$ and $S_0(\rho)$ respectively.  Here, $\rho$ is the areal density of dislocations or, equivalently, the total length of dislocation lines per unit volume. The entropy $S_0(\rho)$ is computed by counting the number of arrangements of dislocations at fixed values of $U_0$ and $\rho$.  

The dislocations are driven by the applied stress to undergo motions that are chaotic on deformation time scales; that is, they explore  statistically significant parts of their configuration space.  According to Gibbs, this configurational subsystem must  maximize its entropy; that is, it finds a state of maximum probability.  It does this at a value of the energy $U_0$ that is determined by the balance between the input power and the rate at which energy is dissipated into the kinetic-vibrational subsystem, which serves here as the thermal reservoir.  The method of Lagrange multipliers tells us to find this most probable state by maximizing the function $S_0 - X U_0$, and then finding the value of the multiplier $X$ for which $U_0$ has the desired value.  Define $X$ to be proportional to the inverse of an effective temperature $T_{e\!f\!f}$, i.e. $1/X = k_B T_{e\!f\!f}\equiv\chi$.   Thus, the system finds a minimum of the free energy 
\begin{equation}
\label{Fdef}
F_0 = U_0 - \chi\,S_0, 
\end{equation}
which, for the moment, is simply a function of $\rho$.  

We already can draw some interesting conclusions.  Note first that minimizing $F_0$ in Eq.~(\ref{Fdef}) determines the steady-state dislocation density, say  $\rho_{0}$, as a function of the steady-state effective temperature, say  $ \chi_{0}$. In the simplest approximation, $U_0 = A\,e_D\, \rho$, where $e_D$ is a characteristic energy of a dislocation of length $L$.  As in LBL, I omit the conventional logarithmic correction for elastic energy primarily because it muddies the algebra unnecessarily, but also because I am not sure it is correct for present purposes. The elastic interactions between dislocations could induce screening correlations in ways not properly described by the logarithmic approximation.  The following arguments will be clearest, and the agreement with experiment will not be impaired, if we simply assume that the single-dislocation energy $e_D$ already includes elastic contributions.  

Similarly, we can estimate the $\rho$ dependence of the entropy $S_0$ by dividing the  area $A$ into elementary squares of area $b^2$, where $b$ is an atomic length scale, roughly the length of the Burgers vector, and then counting the number of ways in which we can distribute $\rho\,A$ line-like dislocations, oriented  perpendicular to the plane, among those squares. The result has the familiar form $S_0 = -A\,\rho\,\ln(b^2\,\rho) + A\,\rho$. Thus, minimizing $F_0$ with respect to $\rho$ produces the usual Boltzmann formula, 
\begin{equation}
\label{rho0}
\rho_{0} = {1\over b^2}\,e^{-\,e_D/\chi_0}. 
\end{equation}
We already see that an appreciable density of dislocations requires a value of $\chi_{0}$ that is comparable to $e_D$, which is enormously larger than the ambient thermal energy $k_B\,T$.   

Next, note that $\chi$ is a measure of the configurational disorder in the material, in direct analogy to the way in which the ordinary temperature determines the strength of low-energy fluctuations.  As such, $\chi_{0}$ must be a function primarily of the plastic shear rate $\dot\epsilon^{pl}$, which we can think of as the rate at which the system is being ``stirred,'' i.e. the rate at which the atoms are being caused to undergo rearrangements.  If this shear rate is so slow that the system has time to relax between rearrangement events, then the steady state of disorder is determined only by the number of atomic rearrangements that have occurred and not by the rate at which they occurred.  That is, $\chi_{0}$ must be some nonzero constant below a characteristic shear rate whose value is of the order of atomic frequencies, i.e. very large.  It follows that $\rho_0$ is also nearly independent of shear rate for normally  slow, steady-state deformations.  

In Section \ref{Depin}, I argue that the driving stress is determined primarily by the density of dislocation entanglements.  Thus,  a constant  $\rho_0$  means a nearly (but not quite) constant steady-state stress over a wide range of low to moderate shear rates.  This is what is observed experimentally. As shown in LBL and also here in Sec.\ref{Depin} and in Fig.~\ref{Rev-Fig-1}, the steady-state stress for room temperature copper increases by less than a factor of $2$ between strain rates of $10^{-3}\,s^{-1}$ and $10^{8}\,s^{-1}$.  In LBL, we also showed that the steady-state value of $\chi$ becomes noticeably strain-rate dependent above rates of the order of $10^{9}\,s^{-1}$; and we  were able to understand the strong-shock results at the higher strain rates by using a glass-theory analogy to write a simple relation between $\chi$ and strain rates up to about $10^{12}\,s^{-1}$. In this paper, however, I want to focus only on the slower deformations because they can be understood in much simpler, more fundamental ways.  

Nevertheless, there is an aspect of the transition from constant to strain-rate dependent values of $\chi$ that illustrates the power of the effective-temperature concept.  In LBL, we suggested that this transition might have a universal geometric interpretation, roughly analogous to the Lindemann criterion according to which crystals melt when thermal vibration amplitudes are of the order of a tenth of the lattice spacing.  In that spirit, we suggested that the value of $\chi_0/e_D$ might correspond to a dislocation density of the order of $(10\,b)^{-2}$, i.e. $e_D/\chi_0 \sim 2\,\ln(10)\sim 4$.  The resulting value $\chi_0/e_D \sim 0.25$ is remarkably close to what we find in our analyses of a wide range of experimental data. 

\section{Depinning Model}
\label{Depin}

The depinning model is literally a mathematical description of the entangled dislocations that Cottrell and Nabarro compared to a ``bird's nest'' because of its supposedly intractable complexity.\cite{COTTRELL-02} In fact, it is that complexity that allows a simple statistical analysis to be accurate.  In this model, the dislocations are immobilized by being pinned to each other.  These pinning interactions can be broken infrequently by ordinary thermal fluctuations.  When a pin is broken, the unpinned segment of a dislocation line moves -- in effect, instantaneously -- to a nearby pinning site. There is no useful distinction between stored and mobile dislocations in this picture; all segments of all the dislocations are immobile (i.e. pinned) almost all the time. Nor is there any role to be played by Peierlsian drag forces.  

The depinning theory starts with Orowan's relation between the plastic strain rate $\dot\epsilon^{pl}$, the dislocation density $\rho$, and the average dislocation velocity $v$:
\begin{equation}
\dot\epsilon^{pl}= \rho\,b\,v,
\end{equation}
where $b$, as before, is the magnitude of the Burgers vector, roughly a lattice spacing.  If a depinned dislocation segment moves a distance of about $\ell \equiv 1/\sqrt{\rho}$ between pinning sites, then $v \sim \ell/\tau_P$, where $1/\tau_P$ is a thermally activated depinning rate given by
\begin{equation}
{1\over \tau_P} = {1\over \tau_0}\,e^{- U_P(\sigma)/k_B T}.
\end{equation}  
Here, $\tau_0$ is a microscopic time of the order of $10^{-12}$ s, and $U_P(\sigma)$ is the activation barrier.  

$U_P(\sigma)$ must be a decreasing function of the stress $\sigma$. For dimensional reasons, $\sigma$ should be expressed in units of some physically relevant stress, which we can identify as the Taylor stress $\sigma_T$ for the following reason.  Suppose that a pinned pair of dislocations must be separated by a distance $b'$ in order to break the bond between them. If these dislocations remain pinned to other dislocations at distances $\ell$,  then this displacement is equivalent to a strain of order $b'/\ell= b'\sqrt{\rho}$ and a corresponding stress of order $\mu\,b'\,\sqrt{\rho}$, where $\mu$ is the shear modulus.   Thus  
\begin{equation}
\label{sigmaT}
\sigma_T(\rho) = \mu\,{b'\over \ell} \equiv \mu_T\,b\,\sqrt{\rho}.
\end{equation}
This is the Taylor stress with $\mu_T = (b'/b)\,\mu$, rederived here by an argument not very different from the one that Taylor used in his 1934 paper.\cite{TAYLOR-34}   As in LBL, we then write 
\begin{equation}
\label{UP}
U_P(\sigma) = k_B\,T_P\,e^{- \sigma/\sigma_T(\rho)},
\end{equation}
where $k_B\,T_P$ is the pinning energy at zero stress. The exponential function in Eq.~(\ref{UP}) has no special significance; in all applications so far, its argument $\sigma/\sigma_T$ varies by no more than a factor of two or three. Note, however, that $\mu_T$ now may contain dimensionless factors that have been suppressed by this choice of stress dependence. Note also that $\sigma$ denotes only the magnitude of the stress in this formula, because this part of the analysis determines only the scalar time scale $\tau_P$.  Directional information will appear in other parts of the stress-strain relations when stresses and strains become tensors.  

The resulting formula for the strain rate $\dot\epsilon^{pl}$ is
\begin{equation}
\label{qdef}
\dot\epsilon^{pl} = {b\over\tau_0}\,\sqrt{\rho} \,\exp\,\Biggl[-\,{T_P\over T}\,e^{-\sigma/\sigma_T}\Biggr].
\end{equation}
Now solve Eq.(\ref{qdef}) for $\sigma$ as a function of $\rho$, $\dot\epsilon^{pl}$ and  $T$:
\begin{equation}
\label{sigmadef}
\sigma = \sigma_T(\rho)\,\,\nu(\rho,\dot\epsilon^{pl},T),
\end{equation} 
where 
\begin{equation}
\label{nudef}
\nu(\rho,\dot\epsilon^{pl},T) = \ln\Bigl({T_P\over T}\Bigr) - \ln\Biggl[\ln\Bigl({b\sqrt{\rho}\over \dot\epsilon^{pl}\tau_0}\Bigr)\Biggr] .
\end{equation}
The quantity $\nu(\rho,\dot\epsilon^{pl},T)$ is a very slowly varying function of its arguments, consistent with the observation that the Taylor stress is a good approximation to the true stress in most circumstances, and also with the observation that the steady-state stress is generally a very slow function of the strain rate. 

\begin{figure}[h]
\centering \epsfig{width=.5\textwidth,file=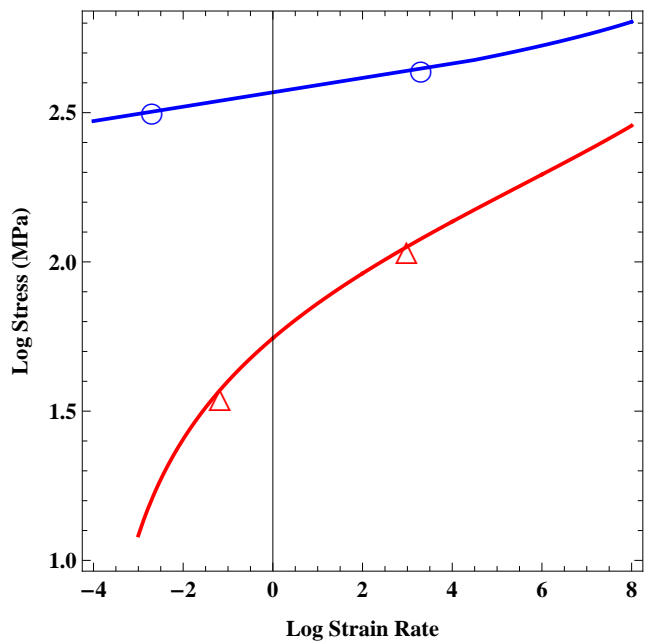} \caption{(Color online) Log-Log plot of the steady-state stresses as functions of strain rate for copper at temperatures $T = 1173\,K$ (lower curve) and $300\,K$ (upper curve).  Parameters are $T_P = 40,800\,K$, $\tau_0 = 10^{-12}\,s^{-1}$, and $\chi_0/e_D \cong 0.25$. The values of $\mu_T$ are $1343\,MPa$ and $1600\,MPa$ for the higher and lower temperatures respectively. The four data points are taken from the stress-strain curves shown in Figs. \ref{Rev-Fig-2} and  \ref{Rev-Fig-3}}\label{Rev-Fig-1}
\end{figure}

This extremely weak rate dependence of the steady-state stress is a remarkable feature of polycrystalline plasticity.  So far as I know, however, it has not been emphasized in the dislocation-theory literature, nor has there been any attempt to explain it other than to include the steady-state stress as a fitting parameter, analogous to the yield stress, in empirical formulas.  In Fig.~\ref{Rev-Fig-1}, I have reproduced some results from LBL showing the steady-state stress as a function of strain rate for copper, for two very different temperatures $T = 1173\,K$ and $300\,K$, over twelve decades of strain rate.  In plotting these curves, I have used Eqs.~(\ref{sigmadef}) and (\ref{nudef}) with parameters shown in the figure caption. The two points marked on each curve are taken from the strain-hardening data shown in Figs. \ref{Rev-Fig-2} and  \ref{Rev-Fig-3}.  The fact that the behavior shown in Fig.~\ref{Rev-Fig-1}, spanning such a wide range of temperatures and strain rates, can be reproduced accurately by the thermodynamic dislocation theory with just a few  physics-based parameters seems to me to be strong support for the basic validity of this theory. 

\section{Nonequilibrium Equations of Motion}
\label{EOM}

We now must move away from the steady-state assumption in order to study time dependent phenomena such as yielding and strain hardening.  This means that we need a careful treatment of the nonequilibrium aspects of these situations, especially a description of the flow of energy and entropy through both the configurational and kinetic-vibrational subsystems.  The following discussion is based on my presentation in \cite{JSL-15} which, in turn, is based on \cite{BLI-09}.  In the latter reference, look especially at Sec. IV for a discussion of the proper definition of internal state variables.   

It is conceptually simplest to keep the model introduced in Sec. \ref{Teff}.  That is, visualize the system as a thin slab of area $A$, thickness $L$, with spatially uniform but time dependent stresses and strains lying in the plane of the slab.  This geometry directly models experiments in which such a slab undergoes simple shear; but it also can be used to describe compression and torsional tests by rescaling the stress and/or the elastic modulus by dimensionless factors of order unity.  (See \cite{LTL-17}.)

Assume, as in Sec.~\ref{Teff}, that the total internal energy of this system can be written as the sum of configurational and kinetic-vibrational parts:
\begin{equation}
U_{total} = U_C(S_C,\rho) + U_R(S_R).
\end{equation}
Here, $U_C(S_C, \rho)$ is the same configurational energy that was introduced earlier except that now it contains not just dislocation energy but also the energies associated with all the other structural degrees of freedom.  In principle, we can include other state variables among the arguments of $U_C$, e.g. the grain size or the density of stacking faults, etc. if we think that those variables might be time dependent and dynamically relevant.  I will not do that here, but will point to situations where those extra variables might be important.  The function $S_C(U_C,\rho)$ is the total entropy of the configurational subsystem, including all structural degrees of freedom, computed by counting the number of configurations at fixed values of $U_C$ and $\rho$. By definition, the effective temperature is
\begin{equation}
\label{chidef}
\chi = \left({\partial U_C\over\partial S_C}\right)_{\rho}.
\end{equation}

$U_R(S_R)$ is the kinetic-vibrational energy of this system, whose entropy is $S_R$.  For most purposes, this subsystem serves as a thermal reservoir.  Its temperature is the ``ordinary'' thermal temperature
\begin{equation}
k_B T \equiv \theta = {\partial U_R\over\partial S_R}. 
\end{equation}
Both $\chi$ and $\theta$ will need to be described by their own equations of motion.    
  
Next, assume that we can write 
\begin{equation}
\label{UC}
U_C(S_C, \rho) = U_0(\rho) + U_1(S_1),
\end{equation}
and, correspondingly, 
\begin{equation}
\label{SC}
S_C(U_C,\rho)= S_0(\rho) + S_1(U_1),
\end{equation}
where $U_0$ and $S_0$ are, respectively, the energy and entropy of the dislocations as introduced in Sec.~\ref{Teff}, and  $U_1$ and $S_1$ are the energy and the entropy of all the other configurational degrees of freedom. As before,
\begin{equation}
\label{Urho}
U_0(\rho)= A\,\rho\,e_D;~~~~e_D = L\,\gamma_D,
\end{equation}
where $\gamma_D$ is the dislocation energy per unit length;
and  
\begin{equation}
\label{Srho}
S_0(\rho) = -\,A\,\rho\,\ln(b^2 \rho)+ A\,\rho .
\end{equation}

The usual thermodynamic analysis for this system is not trivial.  It goes as follows.  The first law is:
\begin{eqnarray}
\label{firstlaw}
\nonumber
\dot U_{total}&=& V\,\sigma\,\dot\epsilon^{pl} = \dot U_C + \dot U_R\cr\\ &=& \chi\,\dot S_C + \left({\partial U_C\over \partial \rho}\right)_{S_C} \dot \rho + \theta\,\dot S_R,
\end{eqnarray}
where $V = LA$ is the volume, $\sigma$ is the shear stress, $\dot\epsilon^{pl}$ is the plastic shear rate, and $ V\,\sigma\,\dot\epsilon^{pl}$ is the mechanical power delivered to the system by external forces. (Variations of the reversible elastic energy cancel out of this equation.) Use Eq.(\ref{firstlaw}) to evaluate $\dot S_C$, and write the second law in the form
\begin{equation}
\label{dotS}
\dot S_C + \dot S_R = {1\over \chi}\,{\cal W} +\left(1 - {\theta\over \chi}\right)\dot S_R \ge 0,
\end{equation}
where
\begin{equation}
\label{W} 
{\cal W} = V\,\sigma\,\dot\epsilon^{pl} - \left({\partial U_C\over \partial \rho}\right)_{S_C}\dot \rho 
\end{equation}
is the difference between the power delivered to the system and the rate at which energy is stored in the form of dislocations.

Equation (\ref{dotS}) is the sum of independent inequalities that, according to an argument originally due to Coleman and Noll \cite{COLEMAN-NOLL-63}, must be satisfied separately.  Non-negativity of the term proportional to $\dot S_R$ implies that the heat flux ${\cal Q}$, defined here to be positive when heat is flowing from the configurational subsystem into the thermal reservoir, is
\begin{equation}
\label{Q}
{\cal Q} = \theta \dot S_R = {\cal K}\,(\chi - \theta),
\end{equation}
where ${\cal K}$ is a non-negative thermal transport coefficient.  

For present purposes, assume that the mechanical power, $V\,\sigma\,\dot \epsilon^{pl}$, is always positive.  Therefore, the remaining inequality is
\begin{equation}
\label{F'neg}
\left({\partial U_C\over \partial \rho}\right)_{S_C}\,\dot \rho \le 0.
\end{equation}
Use Eqs. (\ref{UC}) and (\ref{SC}) to write $U_C = U_0 + U_1(S_C - S_0)$, so that 
\begin{equation}
\left({\partial U_C\over \partial \rho}\right)_{S_C} = {\partial U_0\over \partial \rho} - \chi\,{\partial S_0\over \partial \rho} \equiv {\partial F_0\over \partial \rho};
\end{equation}
where
\begin{equation}
\label{Fdef1}
 F_0(\rho) \equiv U_0(\rho) - \chi\,S_0(\rho)
\end{equation}
is the same $\rho$-dependent free energy that was defined in Eq.(\ref{Fdef}).

Equation (\ref{F'neg}) is satisfied by writing an equation of motion for $\rho$ in the form
\begin{equation}
\label{dotrho}
\dot \rho = -{\cal M}\,{\partial F_0\over \partial \rho},
\end{equation}
where ${\cal M}$ is a non-negative rate factor.  As before, Eqs.(\ref{Urho}) and (\ref{Srho}) imply that $\partial F_0/\partial \rho = 0$ when
\begin{equation}
\label{rho01}
\rho = \rho_0(\chi) = {1\over b^2}\,e^{-\,e_D/\chi}, 
\end{equation}
which is the same Boltzmann formula that appeared in Eq.~(\ref{rho0}) except that Eq.~(\ref{rho01}) is valid for non-steady-state values of $\chi$.  

It is simplest to rewrite Eq.(\ref{dotrho}) in the linearized form
\begin{equation}
\label{dotrho2}
\dot \rho = \tilde {\cal M}\,\left[1- {\rho \over \rho_0(\chi)}\right].
\end{equation} 
The factor $\tilde {\cal M}$ must be proportional to the power per unit volume $\sigma\,\dot \epsilon^{pl}$, which is the only relevant rate in the problem, and which has the dimensions of energy per unit volume per unit time. The left-hand side of Eq.(\ref{dotrho2}) has the dimensions of length (of dislocations) per unit volume per unit time.  Thus, writing this equation in the form
\begin{equation}
\label{dotrho3}
\dot \rho = \kappa_{\rho}\,{\sigma\,\dot \epsilon^{pl}\over \gamma_D}\,\left[1- {\rho \over \rho_0(\chi)}\right]
\end{equation}
is dimensionally correct and identifies the dimensionless factor $\kappa_{\rho}$ as the fraction of the input power that is converted into dislocations.  The second term on the right-hand side of Eq.(\ref{dotrho3}) then can be interpreted as the rate at which dislocations are annihilated as required by the second law. With this detailed-balance argument, there is no need to model specific annihilation mechanisms.  

Having derived an equation of motion for $\rho$, return now to Eq.(\ref{firstlaw}) and rewrite this first-law equation in a form suitable for deriving an equation of motion for $\chi$:
\begin{equation}
\label{firstlaw2}
 \chi\,\dot S_C = V\,\sigma\,\dot \epsilon^{pl} - \left({\partial U_C\over \partial \rho}\right)_{S_C}\dot \rho -{\cal Q}.
\end{equation}
Use the decompositions in Eqs.(\ref{UC}) and (\ref{SC}) to write the left-hand side as
\begin{equation}
\chi\,\dot S_C = \chi\, {\partial S_1\over \partial \chi}\, \dot \chi + \chi {\partial S_0\over \partial \rho}\,\dot \rho \equiv V\,c_{e\!f\!f}\,\dot \chi + \chi {\partial S_0\over \partial \rho}\,\dot \rho,
\end{equation}
which defines the effective heat capacity $V\,c_{e\!f\!f} = \chi\,\partial S_1/\partial \chi$.  Next, make a similar expansion of the right-hand side of Eq.(\ref{firstlaw2}), and note that the term proportional to $\partial S_0/\partial \rho$ cancels out, leaving
\begin{equation}
\label{firstlaw3}
V\,c_{e\!f\!f}\,\dot \chi = V\,\sigma\,\dot \epsilon^{pl}  - {\partial U_0\over \partial \rho}\,\dot \rho -{\cal Q}.
\end{equation}

According to the analysis in Sec.~\ref{Teff}, $\chi$ is comparable to the mesoscopically large energy $e_D$, so that $\chi \gg \theta$ and ${\cal Q} \approx {\cal K}\,\chi$ in Eq.(\ref{Q}). We also know from Sec.~\ref{Teff} that the steady-state value of $\chi$, except at extremely high strain rates, is the constant $\chi_0$.  Thus we can write ${\cal K} = V\,\sigma\,\dot \epsilon^{pl}/\chi_0$, so that, with Eq.(\ref{Urho}), the equation of motion for $\chi$ becomes
\begin{equation}
\label{firstlaw4}
c_{e\!f\!f}\,\dot \chi = \sigma\,\dot \epsilon^{pl}\,\left[1 - {\chi\over \chi_0}\right]  - \gamma_D\,\dot \rho .
\end{equation}
The last term in this equation accounts for energy storage in the form of dislocations.  

The corresponding equation of motion for the kinetic-vibrational temperature $\theta = k_B T$ is
\begin{equation}
\label{thetadot}
\Bigl({c_p\,\rho_d\over k_B}\Bigr)\,\dot\theta = \beta\,\sigma\,\dot \epsilon^{pl}- {\cal K}_2\,(\theta - \theta_0),
\end{equation}
where $c_p$ is the thermal heat capacity per unit mass, $\rho_d$ is the mass density, and $0 < \beta < 1$ is a dimensionless constant known as the Taylor-Quinney factor that determines what fraction of the input power is converted directly to kinetic-vibrational heat.  ${\cal K}_2$ is a thermal transport coefficient and $\theta_0 = k_B\,T_0$ is the ambient temperature.  

Finally, we must write an equation of motion for the stress $\sigma$.  This is not a trivial issue because, to interpret experimental data, we need to describe both elastic and plastic deformations and yielding transitions from one to the other.  That is, we cannot limit ourselves to the relation between the stress and only the plastic deformation rate as in Eq.~(\ref{sigmadef}).  On the other hand, all of the preceding analysis is based on the rule that only properly defined state variables are allowed as arguments of the internal energy or entropy functions.  A large part of the conventional literature on plasticity violates this rule by using the plastic deformation as a state variable, which is illegal. Plastic deformations (as opposed to deformation rates) can be defined only with respect to fixed reference states; but, by definition, irreversible processes forget their histories of past deformation.  Thus, we are not allowed to use mathematical devices like the popular Kroner-Lee decomposition of elastic and plastic displacements.  So far as I can see, the only reasonable alternative is the ``hypoelastoplastic'' approximation in which the elastic and plastic strain rates are assumed to be simply additive, i.e. the total strain rate is $\dot\epsilon = \dot\epsilon^{el} + \dot\epsilon^{pl}$, and 
\begin{equation}
\label{sigmadot}
\dot\sigma = \mu\,\dot\epsilon^{el} = \mu\,(\dot\epsilon - \dot\epsilon^{pl}),
\end{equation} 
where $\mu$ is the elastic shear modulus. This approximation should be accurate so long as $\mu$ is large and the elastic strains are small.  In other circumstances where the elastic and plastic deformations cannot be disentangled in this way, new kinds of physical approximations will be needed. 

\section{Scaling and Dimensionless Variables}
\label{SH} 

In preparation for using the nonequilibrium equations of motion in interpreting experimental data, it is useful to transform to dimensionless variables and thus to identify the relevant dimensionless physical parameters. All of the systems of interest, for the moment, are undergoing spatially uniform shear at constant (elastic plus plastic) rates $\dot\epsilon$.  Therefore, let $Q = \dot\epsilon\,\tau_0$.  Then, replace the time $t$ by the total strain $\epsilon$, and let $\tau_0\,\partial/\partial t \to Q\,\partial/\partial \epsilon$. (The partial derivatives remind us that these functions eventually will depend on position as well as time.)  Let $q = \dot\epsilon^{pl}\,\tau_0$, $\tilde\rho = b^2\,\rho$, $\tilde\chi = \chi/e_D$, and $\tilde\theta = T/T_P$.  All of these dimensionless quantities are functions of $\epsilon$.

In these scaled variables, the equation of motion for the dislocation density, Eq.~(\ref{dotrho3}), becomes
\begin{equation}
\label{dotrho4}
{\partial\tilde\rho\over \partial\epsilon} = \kappa_{\rho}\,\,{\sigma\,q\over \tilde\gamma_D\, Q}\, \Biggl[1 - {\tilde\rho\over \tilde\rho_0(\tilde\chi)}\Biggr],~~~\tilde\rho_0(\tilde\chi) = e^{- 1/\tilde\chi},
\end{equation}
where $\tilde\gamma_D = \gamma_D/b^2 = e_D/b^2L$. The plastic strain rate defined in Eq.~(\ref{qdef}) is
\begin{equation}
\label{qdef1}
q(\epsilon) = \sqrt{\tilde\rho} \,\exp\,\Bigl[-\,{1\over \tilde\theta}\,e^{-\sigma/\sigma_T(\tilde\rho)}\Bigr]; ~~~~\sigma_T(\tilde\rho) = \mu_T\,\sqrt{\tilde\rho}. 
\end{equation}
Its inverse, Eq.~(\ref{sigmadef}), is
\begin{equation}
\label{sigmadef1}
\sigma = \sigma_T(\tilde\rho)\,\,\tilde\nu(\tilde\rho,q,\tilde\theta),
\end{equation} 
where 
\begin{equation}
\label{nudef1}
\tilde\nu(\tilde\rho,q,\tilde\theta) = -\ln(\tilde\theta) - \ln\Biggl[\ln\Biggl({\sqrt{\tilde\rho}\over q}\Biggr)\Biggr] .
\end{equation}

The equation of motion for the effective temperature, Eq.~(\ref{firstlaw4}), becomes 
\begin{equation}
\label{chidot1}
c_{e\!f\!f}\,{\partial \tilde\chi\over\partial\epsilon} = {\sigma\,q\over Q}\left(1-{\tilde\chi\over \tilde\chi_0}\right) - \gamma_D\,{\partial \tilde\rho\over \partial\epsilon}.
\end{equation}
Similarly, the equation of motion for the scaled, ordinary temperature, Eq.~(\ref{thetadot}), is
\begin{equation}
\label{thetadot1}
{\partial\tilde\theta\over \partial\epsilon} = K(\tilde\theta)\,{\sigma\,q\over Q}  - {K_2\over Q}\,(\tilde\theta - \tilde\theta_0).
\end{equation} 
where $K(\tilde\theta) = \beta\,k_B/ (T_P\,c_p\,\rho_d)$, and $K_2$ is the rescaled thermal transport coefficient.  Finally, the stress equation, Eq.~(\ref{sigmadot}), becomes
\begin{equation}
\label{sigmadot1}
{\partial\sigma\over \partial\epsilon} = \mu\,\Biggl[1 - {q(\epsilon)\over Q}\Biggr].
\end{equation}

\begin{figure}[h]
\centering \epsfig{width=.5\textwidth,file=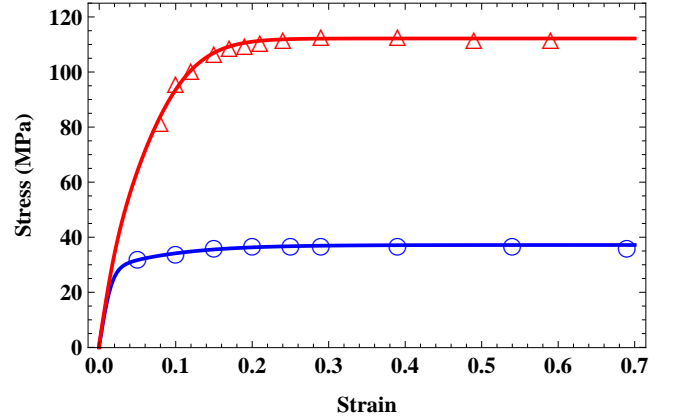} \caption{(Color online) Experimental data and theoretical stress-strain curves for copper at $T=1173\,K$, for strain rates $0.066\,s^{-1}$ (lower blue curve) and $980\,s^{-1}$ (upper red curve).  The parameters used for computing both theoretical curves are $\kappa_1 = 3.1$, $\kappa_2 = 120$, $\chi_0 = 0.25$, and $\mu_T = 1343\,MPa$.  The initial values of $\tilde\rho$  are $10^{-6}$ for both cases; but the initial value of $\tilde\chi$ is $0.22$ for the lower curve and $0.18$ for the upper one.} \label{Rev-Fig-2}
\end{figure}

\begin{figure}[h]
\centering \epsfig{width=.5\textwidth,file=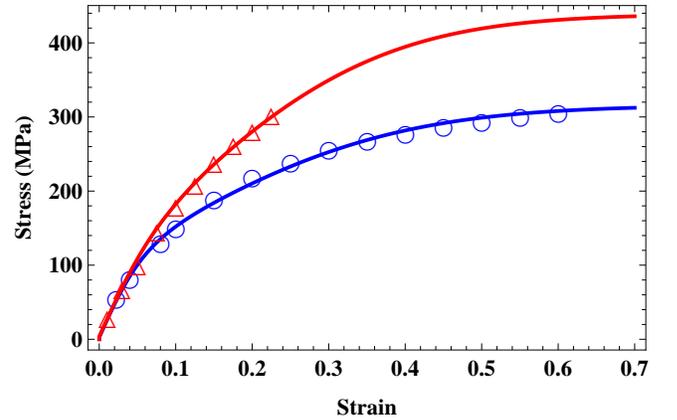} \caption{(Color online) Experimental data and theoretical stress-strain curves for copper at $T=298\,K$, for strain rates $0.002\,s^{-1}$ (lower blue curve) and $2,000\,s^{-1}$ (upper red curve).  The parameters used for computing both theoretical curves are $\kappa_1 = 3.1$, $\kappa_2 = 11.2$, $\chi_0 = 0.25$, and $\mu_T = 1600\,MPa$. The initial values of $\tilde\rho$ and $\tilde\chi$ are $10^{-6}$ and $0.18$ respectively.} \label{Rev-Fig-3}
\end{figure}

The conversion factor $\kappa_{\rho}$ on the right-hand side of Eq.~(\ref{dotrho4}) may usefully be reinterpreted by using an important discovery by Kocks and coworkers.\cite{KOCKS-MECKING-03}  Those investigators found that the onset slope for strain hardening in copper,  $\Theta_0 \equiv (1/\mu)\,(\partial \sigma/\partial \epsilon)_{\rm onset}$, is very nearly constant over a wide range of strain rates and temperatures. Figures \ref{Rev-Fig-2} and  \ref{Rev-Fig-3} show the stress-strain curves for copper on which much of the original LBL theory was based.  In particular, Fig. \ref{Rev-Fig-3} shows two stress-strain curves for room temperature copper at strain rates differing by a factor of $10^6$.  The slopes of these two curves near the origin are indistinguishable from each other, and there is no visible yield stress. 

To understand this behavior, remember that hardening begins when the deformation switches from  elastic to plastic, that is, when $q \cong Q$. In copper, apparently $\tilde\rho \ll \tilde\rho_0$ when this happens, so that Eq.(\ref{dotrho4}) has the form $\partial\tilde\rho/\partial \epsilon \cong \kappa_{\rho}\,\sigma/\tilde\gamma_D \cong \kappa_{\rho}\,\tilde\nu\,\mu_T\,\sqrt{\tilde\rho}/\tilde\gamma_D$.  Assume that we can neglect the extremely slow dependence of $\tilde\nu$ on $\tilde\rho$, so that we can write
\begin{equation}
\Theta_0= {1\over \mu}\,{\partial\sigma\over \partial \epsilon} = {1\over \mu}\,{\partial\sigma\over \partial\tilde\rho}\,{\partial\tilde\rho\over \partial \epsilon} \cong \kappa_{\rho}\,{ \mu_T^2\,\,\tilde\nu^2\over 2\,\mu\,\tilde\gamma_D}.
\end{equation}
Note that $\tilde\rho$ has cancelled out of this expression for $\Theta_0$, and that there is no strain-rate dependence.  Moreover, this result is likely to be independent of temperature because $\tilde\gamma_D$ and the elastic moduli ought to scale thermally in the same ways.  Thus, we have recovered Kocks' result.  

Now, Eq.(\ref{dotrho4}) can be rewritten in the form
\begin{equation}
\label{dotrho5}
{\partial\tilde\rho\over \partial\epsilon} = \kappa_1\,{\sigma\,q\over \tilde\nu^2\,\mu_T\,Q}\, \Biggl[1 - {\tilde\rho\over \tilde\rho_0(\tilde\chi)}\Biggr],
\end{equation}
where the original conversion factor is 
\begin{equation}
\kappa_{\rho} = (2\,\mu\,\tilde\gamma_D/ \tilde\nu^2\,\mu_T^2)\,\Theta_0, 
\end{equation}
and 
\begin{equation}
\label{Theta0}
\kappa_1 = {2\,\mu\over \mu_T} \Theta_0.
\end{equation} 
The quantity $\tilde\gamma_D$ has cancelled out in Eq.~(\ref{dotrho5}), so that the prefactor $\kappa_1$ in Eq.(\ref{Theta0}) is completely determined by directly observable quantities, at least for copper.  

A similarly rewritten version of  Eq.(\ref{chidot1}) is
\begin{eqnarray}
\label{chidot2}
\nonumber
{\partial\,\tilde\chi\over \partial\epsilon}& =& \kappa_2\,{\sigma\,q\over \mu_T\,Q}\,\Biggl[ 1 - {\tilde\chi\over \tilde\chi_0} - {\tilde\gamma_D\,\kappa_1\over \mu_T\,\tilde\nu^2} \Biggl(1 - {\tilde\rho\over \tilde\rho_{ss}(\tilde\chi)}\Biggr)\Biggr]
\cr \\ & \cong & \kappa_2\,{\sigma\,q\over \mu_T\,Q}\,\Biggl( 1 - {\tilde\chi\over \tilde\chi_0} \Biggr).
\end{eqnarray}
Here I have deleted the storage term in the second version of this equation because I have not yet found an experimental situation in which it seems significant. The  dimensionless prefactor is $\kappa_2 = \mu_T/ c_{e\!f\!f}$. Unlike $\kappa_1$, whose value can, in principle, be determined directly from experiment via Eq.(\ref{Theta0}), the coefficient $\kappa_2$  must (so far) be determined by fitting the data. Apart from the relatively unimportant factor $\tilde\nu^{-2}$ in Eq.~(\ref{dotrho5}), the dimensionless prefactors $\kappa_1$ and $\kappa_2$ play comparable roles in these equations.

\section{Comparisons with Experiment} 
\label{EXPT}

The predictions of Eqs.~(\ref{qdef1}), (\ref{nudef1}), (\ref{sigmadot1}), (\ref{dotrho5}), and (\ref{chidot2}) have been tested against a variety of experimental observations. The remarkably accurate agreement between the theory and these observations, while still limited in scope, makes it seem likely to me that this unconventional theory is fundamentally correct.  But there are uncertainties that tell us where to look harder both theoretically and experimentally.  There are also places where the theory implies that conventional interpretations of experimentally observed phenomena are incorrect.  

{\it Copper and the onset of hardening}:  The data points shown in Figs. \ref{Rev-Fig-2} and  \ref{Rev-Fig-3} were taken from \cite{PTW-03,FOLLANSBEE-KOCKS-88,LANL-99}. In the LBL  analysis, we started by using the high-temperature steady-state data in Fig. \ref{Rev-Fig-2}, and the steady-state versions of Eqs.~(\ref{sigmadef}) and (\ref{nudef}), to find  that $T_P \cong 40,800\,K$ and  $\mu/\mu_T \cong 31$, which we assumed to be material-specific parameters, independent of strain rate and temperature.  Note that, in Fig. \ref{Rev-Fig-2}, the steady-state stresses show no signs of thermal softening (unlike the situation for aluminum to be considered next).  

Kocks and Mecking \cite{KOCKS-MECKING-03}, and the graphs shown in Fig.~\ref{Rev-Fig-3}, tell us that $\Theta_0 \cong 1/20$ and thus, from Eq.~(\ref{Theta0}), $\kappa_1 \cong 3.1$.  This is the value of $\kappa_1$ that has been used for all four curves shown here, at all four strain rates and at the two different temperatures.  The story is different for $\kappa_2$, which appears to be strongly temperature dependent but not strongly strain-rate dependent in these cases.  At room temperature, we find $\kappa_2 \cong 11.2$ but, at $T = 1173\,K$, $\kappa_2 \cong 120$. In other words, the effective disorder temperature $\tilde\chi$ moves toward its stationary value more rapidly than does the density of dislocations $\tilde\rho$, and this behavior is amplified at higher temperatures $T$.  This effect can be seen directly in the figures.  In plotting the curves shown here, we did not use the thermal equation of motion, Eq.~(\ref{thetadot1}).  Apparently, the thermal conductivity of copper is large enough that these systems remain close to their nominal ambient temperatures throughout the deformations.  

\begin{figure}[h]
\centering \epsfig{width=.5\textwidth,file=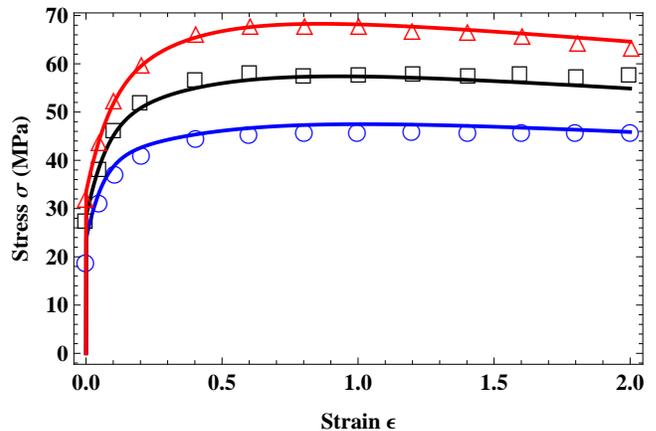} \caption{(Color online) Experimental data and theoretical stress-strain curves for aluminum at $T=573\,K$, for strain rates $0.25\,s^{-1}$ (lower blue curve), $2.5\,s^{-1}$ (middle black curve),  and $25\,s^{-1}$ (upper red curve).  The parameters used for computing all theoretical curves are $\kappa_1 = 0.97$, $\kappa_2 = 12$, $\chi_0 = 0.249$, $T_P = 24,000\,K$, and $\mu_T/\mu = 0.040$.  The initial values of $\tilde\rho$ and $\tilde\chi$ are $0.0035$ and $0.224$ for all cases.} \label{Rev-Fig-4}
\end{figure}

{\it Yield stresses and thermal softening in aluminum}: The situation described in the preceding paragraphs for copper changes markedly in the stress-strain curves for aluminum shown in Fig.~\ref{Rev-Fig-4}. The experimental data shown here is taken from Shi et al \cite{SHIetal-97}; and the theoretical analysis is from \cite{LTL-17}.  These curves are for $T = 573\,K$, for three different strain rates $0.25\,s^{-1}$, $2.5\,s^{-1}$, and $25\,s^{-1}$ .  In  \cite{LTL-17}, we also showed curves for $T = 673\,K$ and $773\,K$ at the same three strain rates, as well as analogous results for a steel alloy. We assumed that all nine of the aluminum samples tested by Shi et al. were prepared identically, and therefore we used a single set of system parameters obtained by a least-squares fit to the data. The values of those parameters are shown in the figure caption.  

\begin{figure}[h]
\centering \epsfig{width=.5\textwidth,file=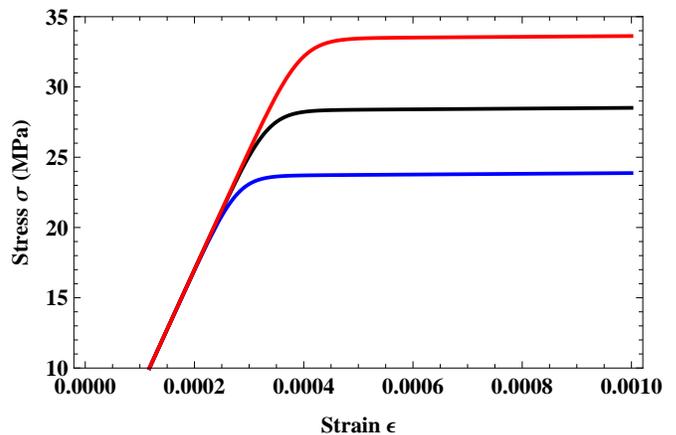} \caption{(Color online) Magnified stress-strain curves for aluminum showing the yielding transitions seen in Fig.~\ref{Rev-Fig-4} at $\epsilon \cong 0$.} \label{Rev-Fig-5}
\end{figure}

The most obvious new feature of these curves is that they exhibit yielding transitions at three different yield stresses for the three different strain rates.  Unlike any other dislocation theories, the thermodynamic theory {\it predicts} these yielding transitions as functions of sample preparation.  The values of the yield stresses and the shapes of the elastic-to-plastic transitions  are determined by the initial values of the dislocation density $\tilde\rho_i$ and the effective temperature $\tilde\chi_i$, which are state variables determined by the prior history of deformation.    

A magnified view of these three theoretical yielding transitions is shown in Fig.~\ref{Rev-Fig-5}.  Here we see a linear elastic stress function with large slope $\mu$, rising rapidly from zero strain, and levelling off sharply but smoothly at three different plastic stresses determined by the three different total strain rates $\dot\epsilon$.  This behavior is governed by the extremely strong stress sensitivity  of $q(\epsilon)$ in Eq.~(\ref{qdef1}), which produces the rapid $\epsilon$-dependence of the stress in Eq.~(\ref{sigmadot1}).  It is important to recognize that this behavior is the converse of the extremely slow strain-rate dependence of the steady-state stress illustrated in Fig.~\ref{Rev-Fig-1}.  The depinning model is playing a central role in explaining two major, apparently disparate features of dislocation plasticity. 

A second way in which these stress-strain curves for aluminum differ from those for copper is that, in Fig.~\ref{Rev-Fig-4}, especially at the largest strain rate, there is clear evidence of thermal softening; the stress decreases at large strains.  This effect is even more visible at the higher temperatures discussed in  \cite{LTL-17}.  To account for it, we have used Eq.~(\ref{thetadot1}), and have set 
\begin{equation}
K(\tilde\theta) = K_0\,\Bigl[1 + c_1\,T_P\,(\tilde\theta - \tilde\theta_1)\Bigr],
\end{equation}
with fitting parameters (the same for all ambient temperatures and strain rates): $K_0 = 7.0 \times 10^{-6}$, $c_1 = 0.0257$, and $T_P\,\tilde\theta_1 = 573\,K$.  This temperature dependence of the thermal conversion factor $K$ produces an interesting nonlinear behavior of the sample temperature $\tilde\theta$. We also used experimentally measured temperature-dependent values for the shear modulus $\mu$.  Note that, in this case, we are using the thermodynamic  dislocation theory to discover thermal properties that we will need, for example, in order to predict fracture toughness or adiabatic shear banding.  In other words, this theory is becoming predictive and potentially falsifiable. 

{\it Hall-Petch effects}: Perhaps the most unexpected result of the thermodynamic theory to date is its unconventional interpretation  \cite{JSL-17} of the Hall-Petch grain-size effects.\cite{ARMSTRONG-HP14}  This result illustrates how the thermodynamic theory can be used as a framework for studying  different physical mechanisms.  Although the depinning model is physically quite specific,  the general thermodynamic equations, e.g. Eqs.~(\ref{dotrho5}) and (\ref{chidot2}) with their dimensionless conversion factors $\kappa_1$ and $\kappa_2$, leave room for exploring other parts of the underlying physics.

In \cite{JSL-17}, I reanalyzed the stress-strain measurements by Meyers et al. \cite{MEYERSetal-95} for room temperature copper with a range of grain diameters, $d = 9.5,\,25,\,117,\,{\rm and}\, 315\,\mu  m$, and for two very different strain rates $10^{-3}\,s^{-1}$ and $3\,\times\,10^{3}\,s^{-1}$.  Because there are no yield stresses in this data, I could use Eq.~(\ref{Theta0}) to evaluate $\kappa_1$ directly as a function of $d$.  The result was that
\begin{equation}
\label{kappa1HP}
\kappa_1^{slow} \cong 2 + {21\over \sqrt{d}}, ~~~~ \kappa_1^{fast} \cong 2 + {60\over \sqrt{d}},
\end{equation}
for the slow and fast cases respectively.   This is exactly the form of the Hall-Petch formulas, which ordinarily describe directly observable yield stresses instead of internal system parameters.  The only difference is that the terms proportional to $d^{-1/2}$ here are larger than the constant terms for all but the largest values of $d$; i.e. the values of $\kappa_1$ seem to be dominated by the grain-size effects.  To test these results, I carried out a series of computational strain-hardening experiments.  That is, I computed the values of $\tilde\rho$ and $\tilde\chi$ at $\epsilon = 0.1$ in the original measurements, and then used those values as initial conditions for computing a new set of stress-strain curves.  The new curves had yield stresses and, as expected, those artificial yield stresses obeyed a Hall-Petch formula of the form $\sigma_y = \sigma_0 + {\rm const.}\,/\sqrt{d}$.  But the constant in this formula was always substantially smaller than $\sigma_0$

\begin{figure}[h]
\centering \epsfig{width=.5\textwidth,file=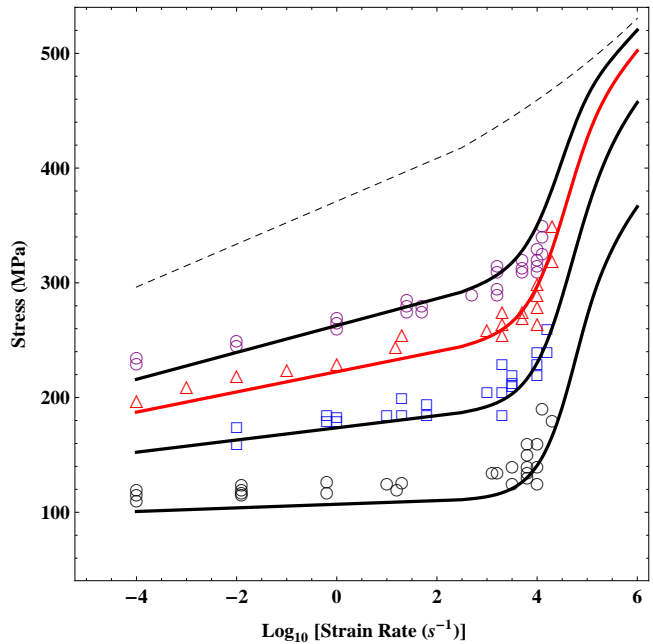} \caption{(Color online) Rate-hardening anomaly.  The four curves, from bottom to top, show stresses as functions of strain rate for four different strains, $\epsilon = 0.05,\,0.10,\,0.15,\,{\rm and}\,0.20$. The dashed curve at the top is the theoretical steady-state prediction.} \label{Rev-Fig-6}
\end{figure}

Apparently, stress concentrations proportional to $d^{-1/2}$ at the edges of the grains amplify the strengths of dislocation sources, resulting in larger dislocation densities and correspondingly larger yield stresses. Thus, the Hall-Petch effect is not caused primarily by dislocation pile-ups at grain boundaries as has conventionally been assumed.  On the contrary, the H-P behavior seems to be caused almost entirely by increased rates of dislocation formation described by the conversion coefficient $\kappa_1$.  

{\it Rate-hardening anomaly:} To end this summary of experimental checks of the thermodynamic dislocation theory, I briefly mention the rate-hardening anomaly that seemed puzzling when not properly understood as a transient phenomenon in the context of a nonequilibrium theory.  The topic again involves the physical content of the conversion coefficient $\kappa_1$. The material is the same room-temperature copper whose stress-strain behavior is shown in Fig.~\ref{Rev-Fig-3}.  The anomaly is the abrupt increase of the flow stress when measured as a function of strain rate at fixed values of the strain in the range $0.05 < \epsilon < 0.20$.\cite{FOLLANSBEE-KOCKS-88,PTW-03} The experimental results and the theory are shown in Fig.~\ref{Rev-Fig-6}.  The theory involves nothing more than including in $\kappa_1$ a rate dependent factor $(1 + Q/Q_0)$, where $Q_0/\tau_0 = 4 \times 10^4\,/s$, corresponding to the location of the anomaly along the strain-rate axis.  I presume that this is a grain-size effect, i.e. that the rate at which dislocations are created at grain boundaries increases with strain rate. 

There are many issues left unresolved by the preceding discussion.  For example, we do not yet have a physics-based model for predicting how the conversion coefficient $\kappa_1$ depends on both the grain size and the strain rate.  More important, as yet we have no physical basis for understanding the values of the coefficient $\kappa_2$ that determines the rate at which the effective temperature of disorder is increased by external driving forces.  Why, for example, does $\kappa_2$ seem to increase as a function of the ordinary temperature?  Does this really happen?  Or are we somehow misunderstanding the onset of hardening in the high-temperature stress-strain curves in Fig.~\ref{Rev-Fig-2}?

\section{Mechanical Properties: Adiabatic Shear Bands}
\label{ASB}

The most important outstanding challenge for a dislocation theory is to explain the mechanical properties of metals and alloys.  What are the mechanisms by which dislocations control brittleness and ductility?  How do they determine fracture toughness and the dynamics of crack propagation?  

The shear-transformation-zone (STZ)\cite{FL-98,FL-11} theory of amorphous plasticity provides only limited guidance.  It seemed  unusual and controversial about a decade ago, primarily because it used an effective temperature in nonequilibrium solid mechanics.  Now, however, it is being used successfully in multidimensional situations to predict, for example, the fracture toughness of metallic glasses.\cite{RYCROFT-EB-12,RYCROFT-EB-16}  But STZ's and dislocations are fundamentally different kinds of flow defects.   STZ formation is a softening mechanism; the larger the density of STZ's, the more easily the system deforms in response to stress.  In contrast, dislocations are hardening defects; the denser they are, the more they are entangled with each other, and the more stress is needed to drive deformation. 

It has been understood for decades that thermal softening is related to failure in polycrystalline solids.  In the present thermodynamic theory, softening occurs via the formula for the plastic strain rate $\dot\epsilon^{pl}$ in Eq.~(\ref{qdef}), where a very small increase in temperature produces a large increase in strain rate.  This strongly nonlinear relation between strain rate, temperature, and stress already has played a central role in explaining sharp yielding transitions and the like.  It must be especially important for understanding the dynamics of crack tips, where the shape of the tip determines the local stress concentration, which controls the local spatially dependent deformation rate, which in turn controls the local rate of heat generation and thus feeds back into the strain-rate formula. But this is a difficult mathematical and computational problem because the stress field is very stiff; it relaxes toward its stationary values much more rapidly  than does the plastic deformation field. This is the problem that Rycroft and colleagues have solved for metallic glasses.\cite{RYCROFT-EB-12,RYCROFT-EB-16}  Addressing it for polycrystalline solids should be high on our list of priorities.

In the absence of a realistic fracture theory, the best I have been able to do so far is to address the problem of adiabatic shear banding (ASB).  There is a large body of literature, extending over more than three decades, devoted to shear banding instabilities in metals and alloys. For example, see \cite{MARCHAND-DUFFY-88, WRIGHT-02, ASL-12}. This subject is important; the banding instability is generally recognized as a principal failure mechanism in rapidly stressed structural materials.  It may often be a precursor to true fracture.   

The ``adiabaticity'' of ASB refers to the idea that these banding instabilities are caused by thermal softening in situations where heat flow is slower than plastic deformation.  A local increase in strain rate produces a local increase in heat generation that, in turn, softens the material and further increases the local strain rate.  The result is a runaway instability if the heat is unable to flow away from the hot spot more quickly than new heat is being generated there. Thus, we are looking at a delicate balance between thermal and mechanical behaviors.  

So far as I know, the best ASB measurements available for analysis are those of  Marchand and Duffy (MD).\cite{MARCHAND-DUFFY-88}  In the next paragraphs, I present an oversimplified analysis of just one part of the MD data.  My purpose is not to provide a detailed theory of the MD results analogous to the theories of strain hardening in copper and aluminum described previously in Sec.~\ref{EXPT}.  As I write, my colleagues K.C. Le and T.M. Tranh are working on that more ambitious project, using all of the MD data for different strain rates and temperatures in an attempt to understand that data in realistic detail.  Here, however, I want to use only the simplest example in order to explore the ways in which the thermodynamic dislocation theory is -- and is not -- able to describe  strongly unstable processes such as ASB.  I already have done this in  \cite{JSL-17}, where I showed how the thermodynamic theory describes ASB in a fictitious material that I called ``pseudocopper.'' Such an analysis becomes more meaningful when carried out in comparison with real observations.  

\begin{figure}[h]
\centering \epsfig{width=.5\textwidth,file=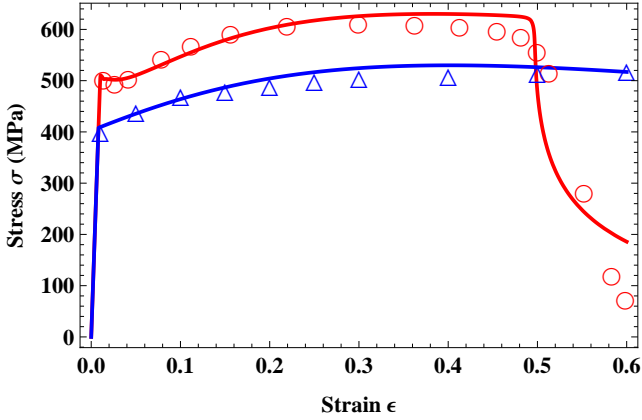} \caption{(Color online) Stress-strain curves for adiabatic shear banding in steel. The experimental data is from Fig.8 of Marchand and Duffy \cite{MARCHAND-DUFFY-88}. The upper curve (red) is for strain rate $\dot\epsilon = 3300\,s^{-1}$, the lower (blue) is for $10^{-4} s^{-1}$.  The system  parameters for both curves are: $T_P = 6 \times 10^5\,K,\,T = 300\,K,\,\mu = 5 \times 10^4\,MPa, \, \mu_T = 1200\,MPa,\,\tilde\chi_0 = 0.25,\, \kappa_1 = 4,\,K_0 = 2 \times 10^{-6},\,{\rm and}\, K_1=K_2= 0$. For the upper curve, $\kappa_2 = 16$, and the initial values of the parameters are $\tilde\rho_i = 0.008,\,\tilde\chi_i = 0.19$.  The inscribed line defect is given by Eq.(\ref{notch}) with $\delta = 0.017,\,y_0 =0.05$.  For the lower curve, $\kappa_2 = 5,\, \tilde\rho_i = 0.007,\,\tilde\chi_i = 0.22$.}   \label{Rev-Fig-7}
\end{figure}

The points shown in Fig.~\ref{Rev-Fig-7}, taken from MD Fig. 8, describe the results of two room-temperature stress-strain experiments.  The red circles are stresses measured at a high strain rate, $\dot\epsilon = 3300\,s^{-1}$, for which the stress drops abruptly at a shear banding instability.  The blue triangles are quasistatic stresses measured at $\dot\epsilon = 10^{-4}\,s^{-1}$, for which the stress rises slowly and smoothly.  These measurements were carried out on thin steel tubes subjected to torsional stresses at constant strain rates.  For present purposes, it suffices to model this system as a thin strip of width $W$ in an $xy$ plane, with the $y$ axis ($-W \le y \le W$) parallel to the tube and the $x$ axis wrapped around a circumference.  The equivalent strip is subject to simple shear parallel to its $x$ axis.  To initiate the instability experimentally, a line defect was inscribed along a circumference, which is equivalent theoretically to making a narrow perturbation along the $x$ axis at $y = 0$, as in Eq.~(\ref{notch}) below.  

The only spatial dependence of the plastic deformation is in the $y$ direction. Thus the equations of motion for the state functions $\tilde\rho$ and $\tilde\chi$, Eqs.~(\ref{dotrho5}) and (\ref{chidot2}), remain unchanged, except that these functions and the  dimensionless strain rate $q$ now depend on $y$ as well as $\epsilon$.  In principle. Eq.~(\ref{thetadot1}) should be modified by adding a diffusion term proportional to $K_1$:
\begin{equation}
\label{thetadot2}
{\partial\tilde\theta\over \partial\epsilon} = K_0\,{\sigma\,q\over Q} + {K_1\over Q}\,{\partial^2 \tilde\theta\over\partial y^2} - {K_2\over Q}\,(\tilde\theta - \tilde\theta_0);
\end{equation} 
but I will not use that term here. All of the terms on the right-hand side of Eq.~(\ref{thetadot2}) may play important roles in a more complete analysis of the MD data. 

In this geometry, the stress remains constant as a function of $y$.  I have found it computionally convenient to enforce the latter constraint by writing  
\begin{equation}
\label{sigmadot2}
{\partial\sigma\over \partial\epsilon} = \mu\,\Biggl[1 - {q(\epsilon,y)\over Q}\Biggr] + M\,{\partial^2 \sigma\over \partial y^2},
\end{equation}
and using a large value of the ``diffusion constant'' $M = 10^5$ to suppress $y$ variations in $\sigma$.  I model the inscribed line defect by choosing the initial value of the effective temperature to be:
\begin{equation}
\label{notch}
\tilde\chi(0,y) = \tilde\chi_i - \delta\,e^{- y^2/2\,y_0^2},
\end{equation}
where $y$ is measured in units of $W$; i.e. I set $W=1$.  

\begin{figure}[h]
\centering \epsfig{width=.5\textwidth,file=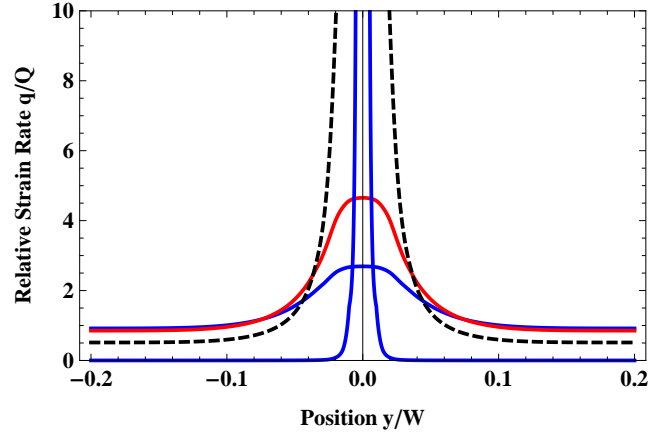} \caption{(Color online) Relative plastic strain rates $q(\epsilon,y)/Q$ at strains $\epsilon = 0.45,\,0.47,\,0.49,$ and $0.497$, for the top stress-strain curve shown in Fig.~\ref{Rev-Fig-7}.  For increasing $\epsilon$, these shear flows are increasingly concentrated in a narrowing band centered at $y=0$. } \label{Rev-Fig-8}
\end{figure}

\begin{figure}[h]
\centering \epsfig{width=.5\textwidth,file=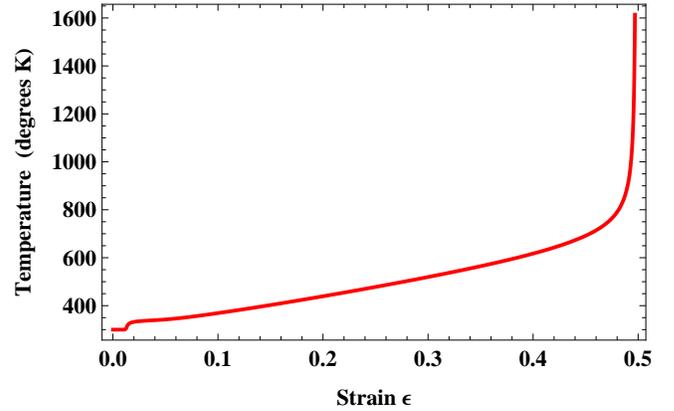} \caption{(Color online) Temperature in degrees K as a function of strain $\epsilon$ at the center of the band, $y = 0$, for the top stress-strain curve shown in Fig.~\ref{Rev-Fig-7}.} \label{Rev-Fig-9}
\end{figure}

The values of the system parameters used for solving these equations are shown in the caption to Fig.~\ref{Rev-Fig-7}.  I have chosen them as follows.  First, I made rough estimates of steady-state stresses for the two curves shown in Fig.~\ref{Rev-Fig-7} and used these, along with Eqs.~(\ref{sigmadef}) and (\ref{nudef}), to estimate $T_P$ and $\mu_T$.  I did the same thing to estimate the initial values $\tilde\rho_i$ at the yield points.  The effective elastic shear modulus $\mu$ can be obtained directly from the initial slopes of these curves, which are visible here unlike in most other yield-stress measurements.  The ratio $\mu/\mu_T \cong 40$ has roughly the same order of magnitude as has been found for other materials.  The theory presented in  \cite{LTL-17} says that this ratio should be a material-specific geometric constant.  

Next, I made a strong adiabatic approximation by setting $K_1 = K_2 = 0$ in Eq.~(\ref{thetadot2}).  Setting $K_1 = 0$ means that there is no intrinsically physical length scale in this model, not even a diffusion length.  The thermal conversion coefficient $K_0$ is primarily responsible for determining the strain at which the shear-banding instability occurs; I have chosen it accordingly.  The values of the conversion factors $\kappa_1$ and $\kappa_2$ were chosen primarily to fit the shapes of the stress-strain curves at intermediate values of $\epsilon$.  As expected, $\kappa_1$ is independent of strain rate.  The different values of $\kappa_2$ look suspicious to me; perhaps this is the result of ignoring some thermal effects and therefore having to use unrealistic values of $\kappa_2$ in order to fit the data.  Finally, I have reduced the initial parameter $\tilde\chi_i$ for the upper curve in order to fit the overshoot seen just above yielding in most of the MD results.  

The abrupt stress drop at $\epsilon  \cong 0.5$ on the fast curve in Fig.~\ref{Rev-Fig-7} indicates the sudden onset of a shear banding instability.  Figures \ref{Rev-Fig-8} and \ref{Rev-Fig-9} show what is happening in more detail . Figure  \ref{Rev-Fig-8}  shows the normalized strain rate $q(\epsilon,y)/Q$ at four different values of $\epsilon$ as the system approaches the transition.  At first, shear localization occurs relatively slowly.  But, at about $\epsilon = 0.49$, this nonlinear process accelerates rapidly.  The plastic strain rate becomes sharply concentrated near $y=0$, causing a sudden increase in the temperature there. The stress decreases uniformly across the system, causing the strain rate to  fall toward zero everywhere except in the increasingly hot band where the runaway instability is occurring.    Fig.~\ref{Rev-Fig-9} shows the temperature at the center of the band.  By $\epsilon \cong 0.49$, this temperature has reached about $800\,K$, which is consistent with the value measured by MD as shown in their Fig. 20.  I find this consistency to be reassuring because I did not adjust any parameters to achieve it.  On the other hand, the extremely rapid collapse of the band shown in Fig.~\ref{Rev-Fig-8}, and the apparent divergence of the temperature above $\epsilon \cong 0.49$ seen in Fig.~\ref{Rev-Fig-9}, tell me  that  both my theory and my numerical analysis have lost their validity at and beyond that point.

The real question, of course, is what this localized meltdown might have to do with ordinary fracture.  It is highly unlikely that thermal singularities of this kind occur at crack tips in real metals and alloys.  Remember that the strain rate in Eq.~(\ref{qdef}) is just as strongly sensitive to changes in the stress as to changes in the temperature.  Thus, a small change in the stress caused by a change in the curvature of a crack tip could produce a runaway instability without melting.  This is why the spatially dependent fracture calculations seem so urgent.

\section{Concluding Questions}

The thermodynamic dislocation theory provides a first-principles framework in which to reexamine our current ideas about the mechanical behaviors of metals and alloys.  Here are some questions that might be parts of that reexamination.

The thermodynamic theory makes no explicit reference to crystal symmetries or distinctions between different kinds of dislocations.  It assumes that all such distinctions are contained in parameters such as the conversion coefficients $\kappa_1$ and $\kappa_2$ and, perhaps, in the dimensionless ratio between $\mu_T$ and $\mu$. Is this correct?  Or do we need some more elaborate change in the theory, for example, different density variables $\tilde\rho$ for different kinds of dislocations?  

In Sec.~\ref{Teff}, I ignored the logarithmic term in the energy that accounts for elastic interactions between dislocations.  Am I making a qualitative mistake by doing this?  The elastic interactions almost certainly are the cause of the cellular dislocation patterns commonly observed in micrographs.  I am fairly sure that the thermodynamic theory can be used to explain those patterns, presumably by generalizing the theory so that $\tilde\rho$ and $\tilde\chi$ become spatially varying order parameters.  The important question is whether these patterns play any role in the dynamics of hardening or failure.  

A more general version of the last question is one that I have asked in almost all of the preceding papers  \cite{LBL-10,JSL-15,JSL-15y,JSL-16,JSL-17,LTL-17}.  In looking at the large range of phenomena that seem to be relevant to polycrystalline plasticity, how can we distinguish between causes and effects?  How can we determine whether an observed structural change such as the appearance of stacking faults or dynamically recrystallized grains (DRX) is a cause of some qualitative change in behavior or simply a side effect of something else that is happening?  In  theoretical language, we need to ask whether the densities of stacking faults or DRX grains or the like are dynamically relevant state variables that need to be included along with $\tilde\rho$ and $\tilde\chi$  in the fundamental equations of motion. 

And finally, I repeat my assertion that the thermodynamic dislocation theory is now well enough developed to be applied to fracture dynamics.    

\begin{acknowledgments}

JSL was supported in part by the U.S. Department of Energy, Office of Basic Energy Sciences, Materials Science and Engineering Division, DE-AC05-00OR-22725, through a subcontract from Oak Ridge National Laboratory.  He thanks K.C. Le for encouragement throughout this project and for ongoing collaboration in some of the computations.  

\end{acknowledgments}

\end{document}